\documentclass{iopart}

\usepackage{iopams}  
\usepackage{graphicx}

\begin{document}

\title[Novel heavy flavor suppression mechanisms in the QGP]
{Novel heavy flavor suppression mechanisms in the QGP}

\author{I Vitev$^\S$, A Adil$^\dag$, H van Hees$^\ddag$}

\address{$^\S$Los Alamos National Laboratory, Theoretical and Physics
Divisions,   Los Alamos, NM 87544, USA \\
$^\dag$Columbia University, Department of Physics, 
New York, NY 10027, USA \\
$^\ddag$Texas A\&M University, Cyclotron Institute, College Station, 
TX 77843, USA }
\ead{ivitev@lanl.gov}

\begin{abstract}
We revisit the question of the measured,  unexpectedly 
large, heavy flavor  suppression, $R_{AA}(p_T) \ll 1$, in 
nucleus-nucleus collisions  at RHIC and compare two new 
theoretical approaches to the $D$- and  $B$-meson quenching. 
In the first model, radiative energy loss, collisional 
energy loss and heavy quark-resonance interactions 
are combined to evaluate the drag and diffusion coefficients 
in the quark-gluon plasma and the mixed phase. These 
are applied in a relativistic Fokker-Planck equation to 
simulate the heavy $c$- and $b$-quark suppression rate and 
elliptic flow $v_2(p_T)$. In the second model, the fragmentation 
probability for heavy quarks and the medium-induced decay
probability for heavy hadrons are derived. These are implemented 
in a set of coupled rate equations that describe the 
attenuation of the observable spectra from the 
collisional dissociation of heavy mesons in the QGP.
An improved description of the non-photonic electron $R_{AA}(p_T)$ 
at RHIC can be obtained. In contrast to previous results, 
the latter approach predicts suppression  of $B$-mesons comparable to 
that  of $D$-mesons at transverse momenta as low as 
$p_T \sim 10$~GeV.
\end{abstract}

\section{Introduction}

The detailed suppression pattern, $R_{AA}(p_T)$, and elliptic flow,
$v_2(p_T)$, of high-transverse-momentum hadrons is an important
experimental signature of the quark-gluon plasma creation in heavy ion
collisions~\cite{Gyulassy:2003mc}. Jet quenching for light mesons, such
as $\pi$, $K$ and $\eta$, at RHIC is well explained by radiative energy
loss calculations~\cite{Vitev:2005he}. It also gives the dominant
contribution to the azimuthal asymmetry of hard
probes~\cite{Gyulassy:2003mc}.  
In contrast, models~\cite{Wicks:2005gt} with a physically reasonable set
of QGP temperatures and densities, predict a QCD heavy-quark energy loss
which is too small compared to the measured suppression of single
non-photonic electrons~\cite{Adare:2006nq,Abelev:2006db}.  Therefore, it
is critical to investigate new interaction mechanisms in the QGP that
may be specific to heavy
flavor~\cite{vanHees:2004gq,vanHees:2005wb,prep,Adil:2006ra}.

\section{Heavy-flavor suppression in a combined transport + 
quenching approach }

Thermalization of heavy quarks in the QGP-heat bath has been recently
studied in the framework of the parton-transport
approach~\cite{vanHees:2004gq,vanHees:2005wb,Molnar:2006ci}.  Large
nuclear suppression and elliptic flow $v_2$ result when employing a
Fokker-Plank equation,
\begin{equation}
\frac{\partial f(\vec{p},t)}{\partial t} = 
\frac{\partial}{\partial p_i} \left[ p_i A(\vec{p},t) + 
\frac{\partial}{\partial p_j}  B_{ij}(\vec{p},t) \right] \; ,
\label{fp}
\end{equation}
solved via an equivalent Langevin simulation. In Eq.~(\ref{fp})
$f(\vec{p},t)$ is the distribution of $c$- and $b$-quarks and
$A(\vec{p},t), B(\vec{p},t)$ are the drag / diffusion coefficients,
respectively. It has been argued that strong coupling between the $c$-
and $b$-quarks and the QGP medium may be generated via quark-resonance
interactions near the QCD-phase transition, $T \sim
T_c$~\cite{vanHees:2004gq}.  It is, therefore, important to study the
interplay between such non-perturbative effects and the
radiative~\cite{Gyulassy:2000er} and collisional~\cite{Wicks:2005gt}
heavy-quark energy loss. We evaluate~\cite{prep} the contribution of
these processes to the drag and diffusion coefficients,
\begin{equation} 
A(\vec{p},t) = \frac{1}{p_i} 
\frac{ \langle  \delta p_i  \rangle}{\delta t} \;, 
 \qquad   
B_{ij}(\vec{p},t) = \frac{1}{2} 
\frac{ \langle  \delta p_i \delta p_j  \rangle} { \delta t} \; ,   
\label{coefs}
\end{equation}
which are then applied in the relativistic Fokker-Planck  equation.

\begin{figure}[!t]
\begin{center}
\includegraphics[width=6.2cm]{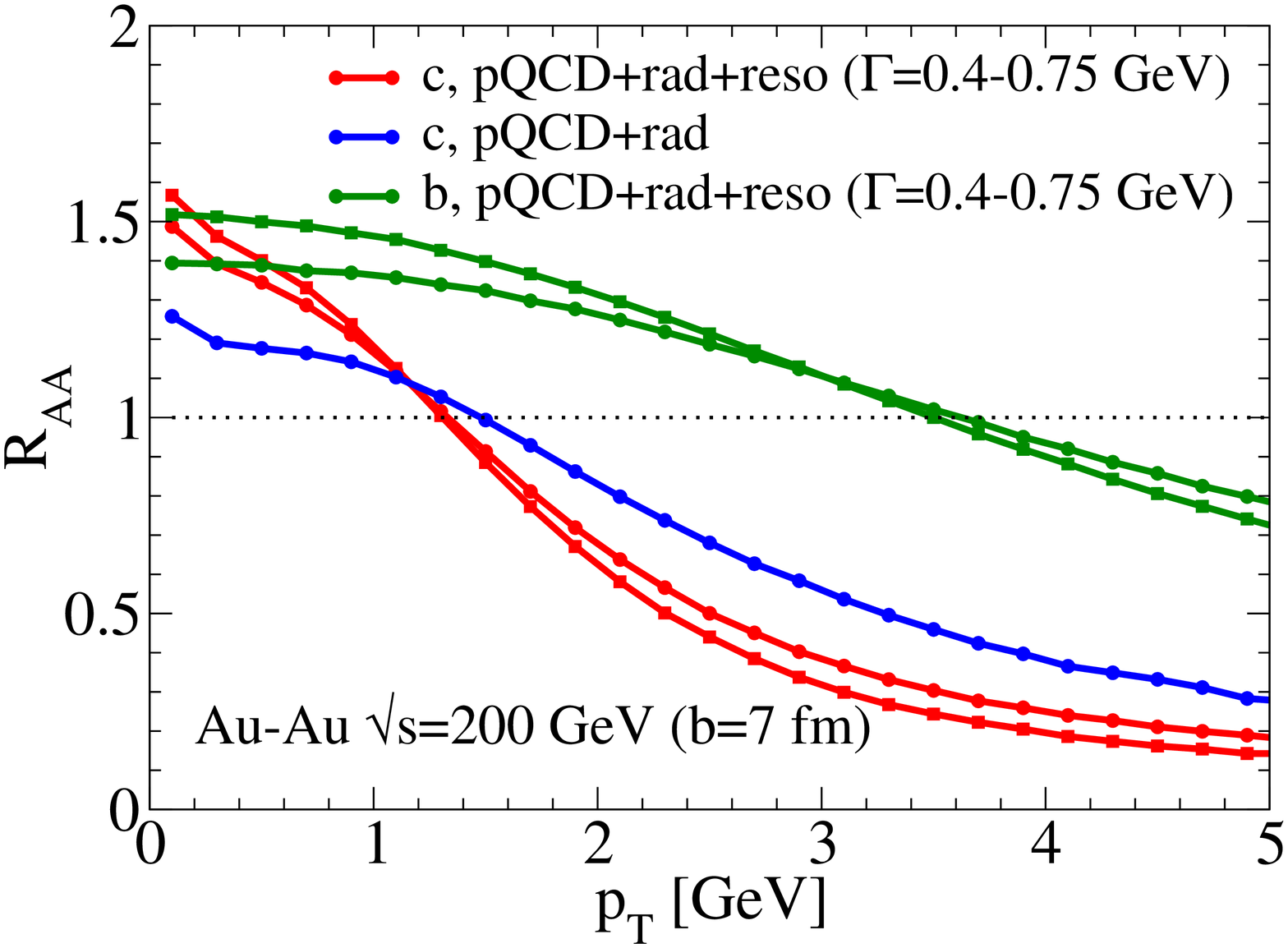}
\hspace*{.2cm} \includegraphics[width=6.2cm]{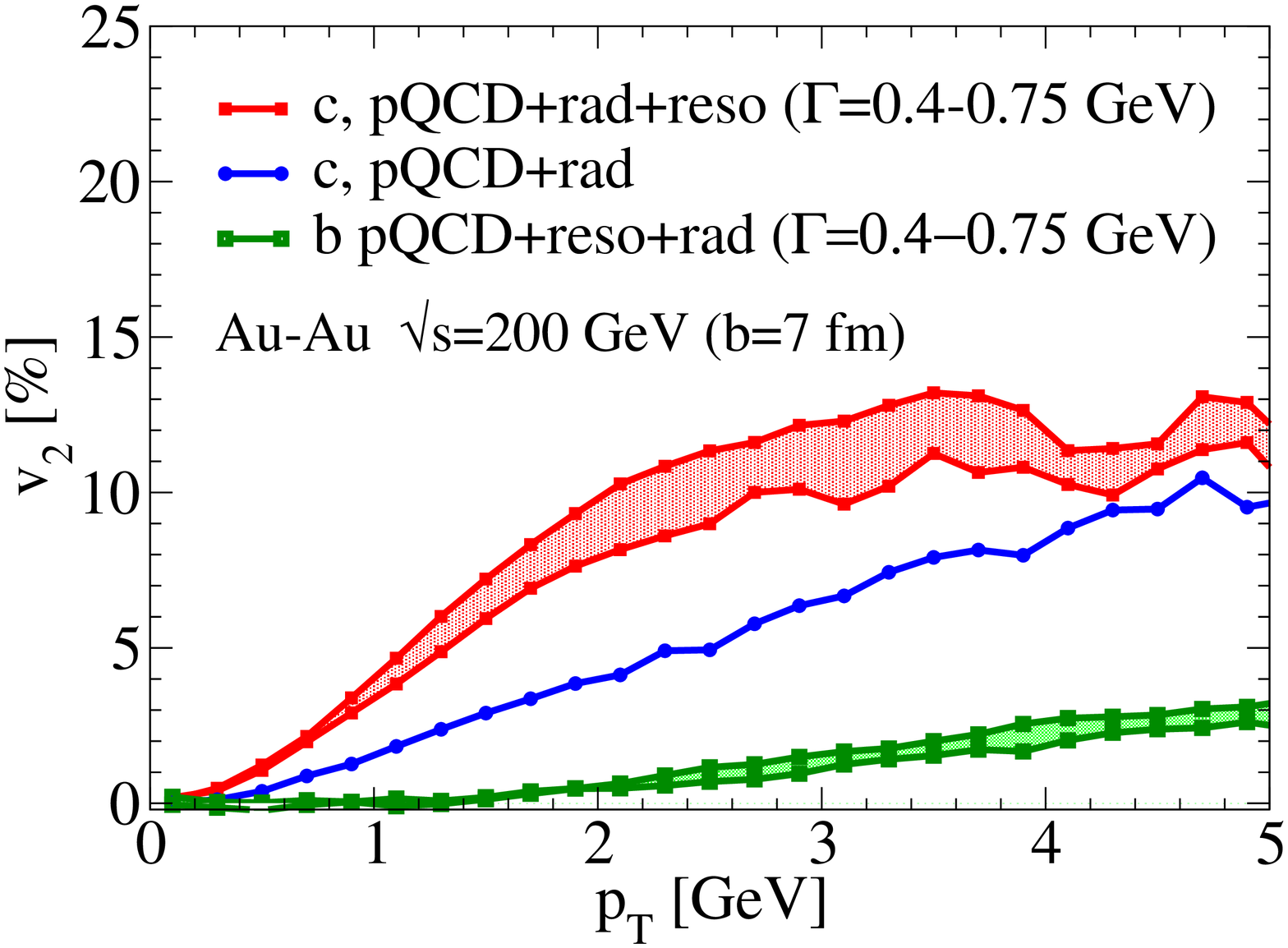}
\end{center}
\vspace*{-0.3cm}
\caption{ Left panel: preliminary results on the nuclear modification  
$R_{AA}(p_T)$ for heavy $c$- and $b$-quarks~\cite{prep}   from 
collisional~\cite{Wicks:2005gt} and 
radiative energy loss~\cite{Gyulassy:2000er} and quark-resonance 
interactions~\cite{vanHees:2004gq,vanHees:2005wb}.
For charm quarks, the PQCD $\Delta E$  contribution is  shown separately.
Right panel: elliptic flow $v_2(p_T)$ for heavy c- and b-quarks 
for the same physics mechanisms~\cite{prep}. }
\label{figure1}
\vspace*{-2mm}
\end{figure}

Results on the $p_T$-dependent suppression pattern
of heavy quarks, $R_{AA}(p_T)$, are shown in the left panel of 
Fig.~\ref{figure1}. Drag coefficients are easily evaluated from 
the fractional momentum loss of heavy quarks, see Eq.~(\ref{coefs}). 
The diffusion coefficients in this preliminary study were constrained 
from the fluctuation-dissipation relation.  We observe that the 
suppression 
of charm quarks can be very large even in minimum-bias 
reactions of large nuclei and $R_{AA}({\rm charm}) \ll 
R_{AA}({\rm bottom})$. The high-$p_T$ azimuthal asymmetry
for minimum-bias Au+Au collisions is shown in  the right 
panel of Fig.~\ref{figure1}.  We note that the generated 
$v_2$ for $b$-quarks is much smaller than that for $c$-quarks.

One of the reasons for the large suppression in our current energy-loss
implementation is that the Einstein fluctuation-dissipation relation
induces minimal Gaussian fluctuations.  These are significantly
different from the ones in the probabilistic treatment of PQCD-energy
loss~\cite{Gyulassy:2003mc,Vitev:2005he,Wicks:2005gt}.  Future Langevin
simulations of $c$- and $b$-quark diffusion should include momentum
fluctuations beyond the Einstein's relation and the decay of the heavy
quark / hadron spectra into $(e^+ + e^-)$ for direct comparison to the
non-photonic electron observables at RHIC~\cite{prep}.

\section{QGP-induced dissociation of heavy mesons}

In the perturbative QCD-factorization approach, the cause of 
the limited  single non-photonic electron quenching is 
identified as the small suppression of $B$-mesons, 
which dominate the  high-$p_T$  $e^+ e^-$ yields.  
Such models assume that the hard jet hadronizes in vacuum,
having fully traversed the region of hot and dense nuclear matter, 
$L_T^{\rm QGP} \leq 6$~fm,  and lost energy via radiative and 
collisional 
processes~\cite{Gyulassy:2003mc,Vitev:2005he,Wicks:2005gt}.
In Ref.~\cite{Adil:2006ra} we examined the validity of this 
assumption for different species of final-state partons and 
decay hadrons.  For a $p_T = 10$~GeV pion at mid-rapidity
$\tau_{\rm form} \approx 25$~fm $\gg L_T^{\rm QGP}$, consistent with 
the jet-quenching assumptions~\cite{Gyulassy:2003mc,Vitev:2005he}.   
In contrast, $B$- and $D$-mesons of the same $p_T$ have 
formation times $\tau_{\rm form} \approx 0.4, 1.6$~fm, 
respectively,  $\ll L_T^{\rm QGP}$.  
Therefore, at the finite $p_T$ range 
accessible at RHIC and LHC a conceptually different approach 
to the description of $D$- and $B$-meson quenching in A+A 
collisions is required, when compared to light hadrons.

\begin{figure}[!t]
\vspace*{+0.cm}
\begin{center}
\includegraphics[width=6.3cm]{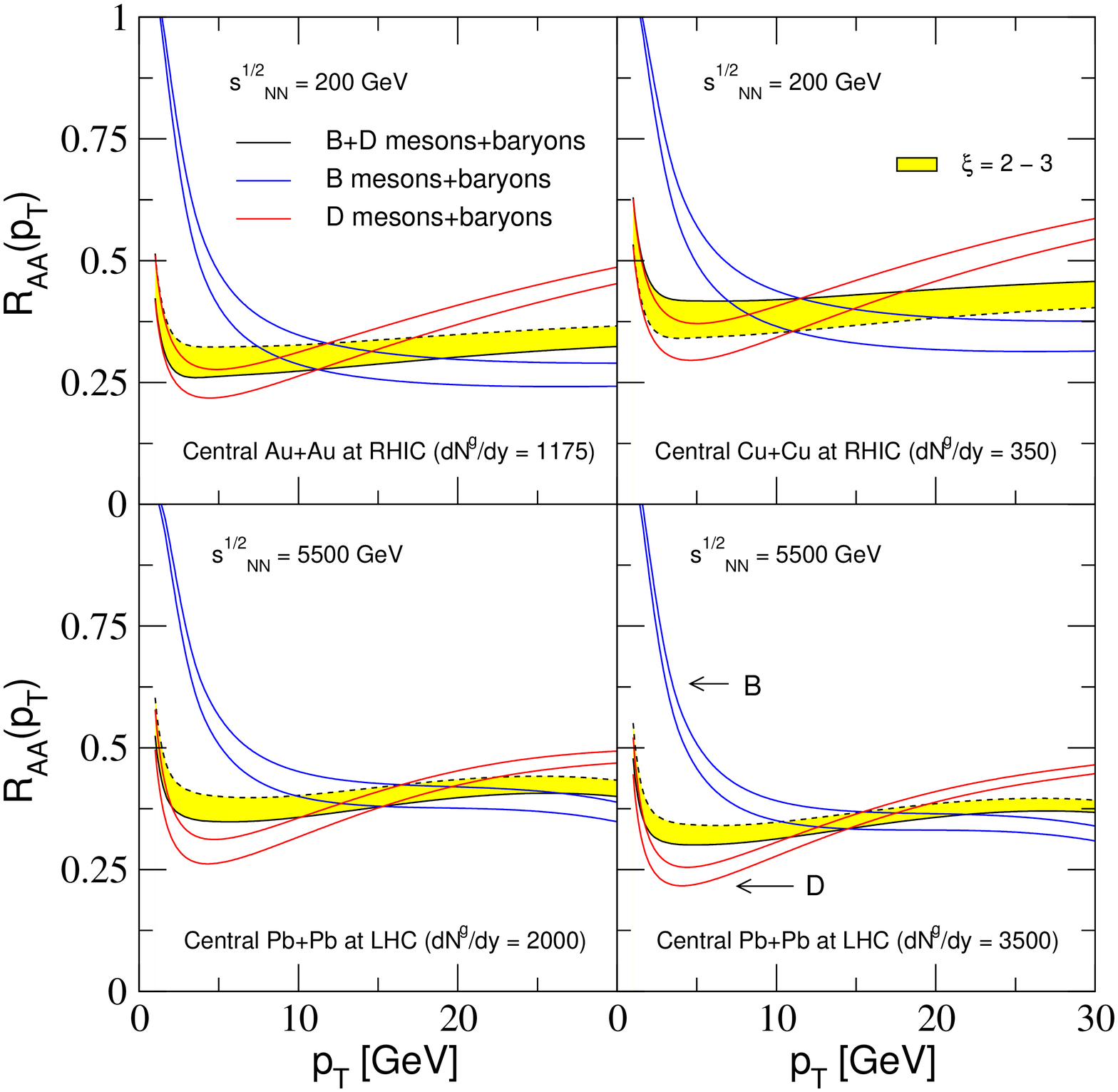}
\hspace*{.2cm}\includegraphics[width=6.3cm]{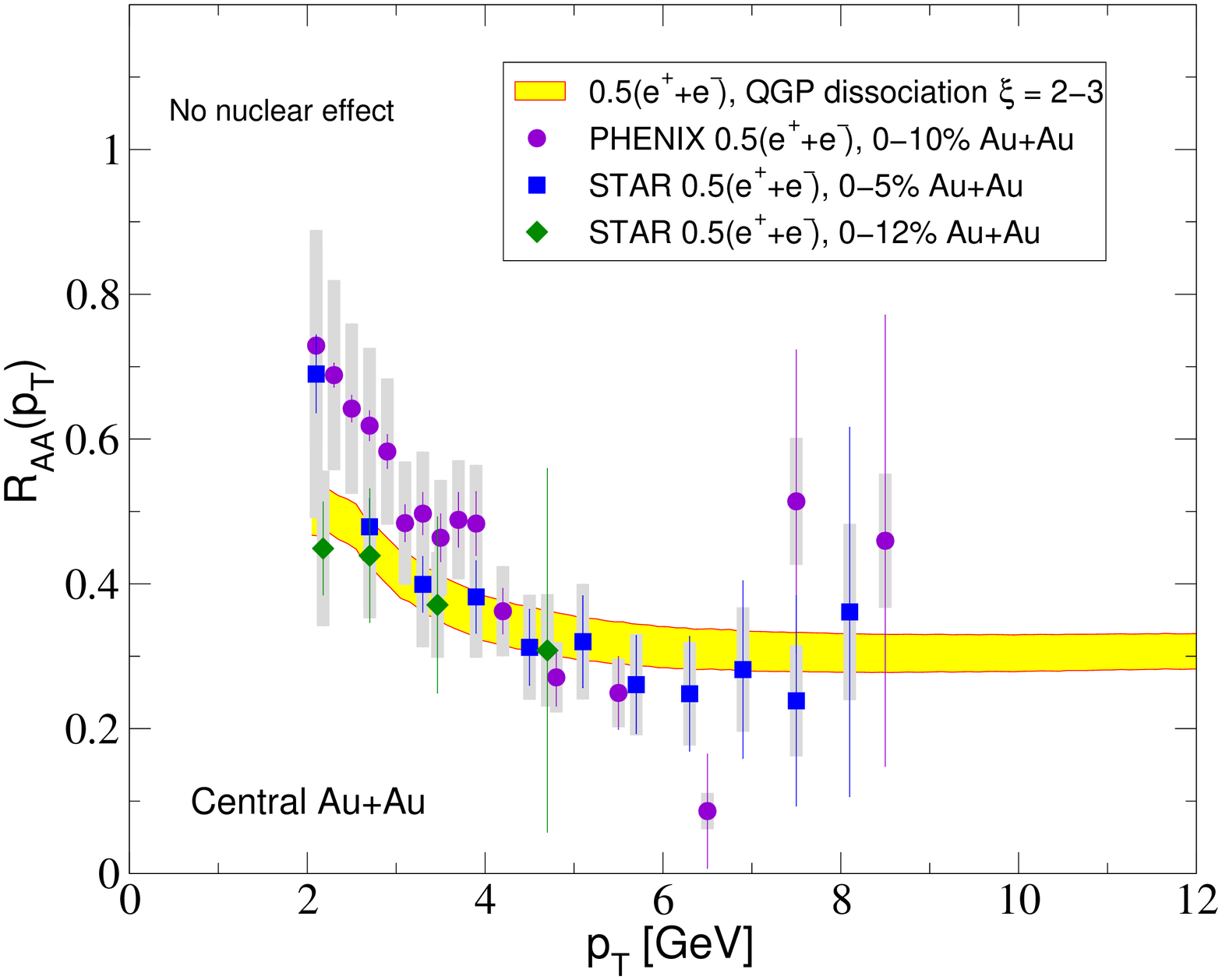}
\end{center}
\vspace*{-0.3cm}
\caption{Left panel: suppressions of $D$- and $B$-meson production via
collisional dissociation in the QGP in central Au+Au and Cu+Cu 
reactions  at RHIC~\cite{Adil:2006ra}
for gluon rapidity densities $dN^g/dy = 1175$ and  
$350$, respectively~\cite{Vitev:2005he}. Right panel:
quenching of the non-photonic electrons from the softened
$D$- and $B$-meson spectra in central  Au+Au 
collisions~\cite{Adil:2006ra}. Data is from  
PHENIX~\cite{Adare:2006nq,ralf} and 
STAR~\cite{Abelev:2006db,alex,zhong}.  } 
\label{figure2}
\vspace{-2mm}
\end{figure}

Motivated by this finding, in the framework of the GLV theory, 
we derive the collisional dissociation probability of heavy 
mesons in the QGP~\cite{Adil:2006ra}:  
\begin{eqnarray}
&&   P_d (\chi\mu^2 \xi) = 
 \left[ 1- P_s (\chi\mu^2 \xi)  \right] \geq 0 \;, 
\quad P_d (\chi\mu^2 \xi = 0) = 0 \;.
\label{sprob}
\end{eqnarray}
In Eq.~(\ref{sprob}) $2 \chi \mu^2 \xi = 2 (\mu^2 L / \lambda) \xi$
is the cumulative 2D transverse momentum squared per parton. The 
dissociation probability also depends on the detailed heavy meson 
light cone wave function. The dynamics of open heavy flavor
production and modification in this model 
is represented by a set of coupled rate  equations that 
describe the competition between $b$- and $c$-quark
fragmentation and $D$- and $B$-meson dissociation~\cite{Adil:2006ra}:  
\begin{eqnarray}
\label{rateq1}
&&  \hspace*{-2.5cm} \partial_t f^{Q}({p}_{T},t) =  
-  \frac{f^{Q}({p}_{T},t)}{\langle \tau_{\rm form}(p_T, t) \rangle} 
  + \, \frac{1}{\langle 
\tau_{\rm diss}(p_T/\bar{x}, t) \rangle}
\int_0^1 dx \,  \frac{1}{x^2} \phi_{Q/H}(x)  
f^{H}({p}_{T}/x,t) \;, \qquad \\
&&  \hspace*{-2.5cm} \partial_t f^{H}({p}_{T},t) =  
-  \frac{f^{H}({p}_{T},t) }{\langle \tau_{\rm diss}(p_T, t) \rangle} 
 +\, \frac{1}{\langle 
\tau_{\rm form}(p_T/\bar{z}, t) \rangle}
\int_0^1 dz \,  \frac{1}{z^2} D_{H/Q}(z) 
 f^{Q}({p}_{T}/z,t) \;. \qquad 
\label{rateq2}
\end{eqnarray}
In Eqs. (\ref{rateq1}) and (\ref{rateq2})  $f^i(p_T,t) =   
d\sigma^i/dy d^2 p_T$. For further details, see~\cite{Adil:2006ra}.

We solve this system of coupled rate equations numerically,
using the same initial soft-gluon rapidity density $dN^g/dy$ as
in the calculation of the $\pi^0$ quenching~\cite{Vitev:2005he}
in central Au+Au and Cu+Cu collisions at RHIC. Our results, 
including a study of a range of anticipated QGP densities at 
the LHC, are shown in the left panel of Fig.~\ref{figure2}. 
Contrary to calculations that emphasize radiative and collisional
heavy quark energy loss~\cite{Wicks:2005gt,vanHees:2005wb,prep},  
QGP-induced  dissociation predicts $B$-meson suppression comparable 
to or larger than that of $D$-mesons at transverse momenta as low as 
$p_T \sim 10$~GeV~\cite{Adil:2006ra}. The heavy meson spectra  
are decayed into electrons $(e^++e^-)$ using the PYTHIA event 
generator.  Our results are shown in the right panel of 
Fig.~\ref{figure2}. The predicted $R_{AA}(p_T)$, which does not 
neglect the large $B$-meson contribution, describes well the most 
recent heavy flavor quenching measurements at 
RHIC~\cite{alex,zhong,ralf}.
We  emphasize that such agreement between theory and experiment 
is not achieved at the cost of neglecting the contribution of 
the $B$-mesons to the non-photonic $e^+ e^-$ spectra.

\section{Conclusions}

In these proceedings, we compared two new theoretical 
approaches~\cite{prep,Adil:2006ra} to open heavy flavor
modification in the QGP. Preliminary  results  on  Langevin
simulations of heavy  quark diffusion, which include radiative
energy loss, collisional energy loss and quark-resonance 
interactions, were shown. While an improved  implementation of 
momentum fluctuations is required for quantitative 
comparison between data~\cite{Adare:2006nq,Abelev:2006db}   
and theory~\cite{prep}, 
we find normal suppression, $R_{AA}(c)\ll  R_{AA}(b)$, 
and elliptic flow, $v_{2}(c)\gg  v_{2}(b)$, hierarchies as 
function of the heavy quark mass.   
Results on  QGP-induced collisional dissociation of 
heavy mesons~\cite{Adil:2006ra} were also shown. 
A good description~\cite{alex,zhong,ralf} of the large 
quenching of the inclusive non-photonic 
electrons~\cite{Adare:2006nq,Abelev:2006db}
is achieved by this model.  A natural consequence of the approach 
developed in Ref.~~\cite{Adil:2006ra}  is 
that $B$-mesons  are attenuated as much as $D$-mesons at 
transverse momenta as  low as $p_T \sim 10$~GeV. 
We conclude that robust experimental determination  of the dominant 
mechanism for in-medium modification of open heavy flavor would require 
direct and separate measurements of the $B$- and $D$-meson  
$R_{AA}$ and  $v_2$ distributions versus $p_T$ and centrality in 
collisions of heavy  nuclei.

\vspace*{0.2cm}\noindent\textbf{Acknowledgment:} This work is
supported by the U.S. Department of Energy under contract no. 
DE-AC52-06NA25396 (IV) and grant no. DE-FG02-93ER40764 (AA)  and  
by the U.S. National Science Foundation under grant no. PHY-0449489 
(HvH).


\vspace*{-.2cm}
\section*{References}
\begin{thebibliography}{20}


\bibitem{Gyulassy:2003mc}
  M.~Gyulassy, I.~Vitev, X.~N.~Wang and B.~W.~Zhang,
  nucl-th/0302077.


\bibitem{Vitev:2005he}
  I.~Vitev,
  Phys.\ Lett.\ B {\bf 639}, 38 (2006), references therein.



\bibitem{Adare:2006nq}
  A.~Adare {\it et al.} [PHENIX Collaboration],
  nucl-ex/0611018.


\bibitem{Abelev:2006db}
  B.~I.~Abelev {\it et al.}  [STAR Collaboration],
  nucl-ex/0607012.


\bibitem{Wicks:2005gt}
  S.~Wicks, W.~Horowitz, M.~Djordjevic and M.~Gyulassy,
  nucl-th/0512076.



\bibitem{vanHees:2004gq}
  H.~van Hees and R.~Rapp,
  Phys.\ Rev.\ C {\bf 71}, 034907 (2005).


      \bibitem{vanHees:2005wb}
        H.~van Hees, V.~Greco and R.~Rapp,
        Phys.\ Rev.\ C {\bf 73}, 034913 (2006).




\bibitem{Molnar:2006ci}
  D.~Molnar,
  nucl-th/0608069, references therein.




\bibitem{Gyulassy:2000er}
  M.~Gyulassy, P.~Levai and I.~Vitev,
  Nucl.\ Phys.\ B {\bf 594}, 371 (2001).



\bibitem{prep}
  H.~van Hees, I. Vitev and R.~Rapp,
  in preparation.


\bibitem{Adil:2006ra}
  A.~Adil and I.~Vitev,
  hep-ph/0611109.



\bibitem{alex}
  A. Suaide, these proceedings, J. Phys. G to be
  published 


\bibitem{zhong}
  C. Zhong [STAR Collaboration], these proceedings, 
  J. Phys. G to be published 

\bibitem{ralf}
  R. Averbeck, these proceedings, J. Phys. G to be
  published 




\end{thebibliography}
\end{document}